\documentclass[iop,apjl]{emulateapj_1}
\usepackage{amsmath,bm,mathrsfs}
\usepackage[colorlinks,citecolor=blue,bookmarks=false,hypertexnames=true]{hyperref} 

\newcommand{\threej}[6]{\left( \begin{array}{ccc}
    #1 & #2 & #3 \\ #4 & #5 & #6
    \end{array} \right)}
\newcommand{\sixj}[6]{\left\{ \begin{array}{ccc}
    #1 & #2 & #3 \\ #4 & #5 & #6
\end{array} \right\}}

\newcommand{\apx}[1]{^{\mbox{\tiny(#1)}}}

\begin{document}

\title{\textsc{Experimental Testing of Scattering Polarization Models}}

\author{W.~Li$^{1,}$\altaffilmark{$^\ast$}}
\author{R.~Casini$^1$}
\author{S.~Tomczyk$^1$}
\author{E.~Landi Degl'Innocenti$^{2,}$\altaffilmark{$^\dagger$}}
\author{B.~Marsell$^3$}

\affiliation{$^1$High Altitude Observatory,
National Center for Atmospheric Research\altaffilmark{$^\ddagger$},
P.O.~Box 3000, Boulder, CO 80307-3000, U.S.A.}
\affiliation{$^2$Dip.\ di Astronomia e Scienza dello Spazio,
Universit\`a di Firenze, L.go E.~Fermi 2, I-50125 Firenze, Italy}
\affiliation{$^3$Stetson University, 421 N Woodland Blvd, DeLand, FL 32723, U.S.A.}

\altaffiltext{$^\ast$}{Corresponding author: wenxianli10@yahoo.com}
\altaffiltext{$^\dagger$}{Egidio Landi Degl'Innocenti was one of the promoters of the experiment, and a principal contributor to its interpretation and to the drafting of this letter. Sadly, he passed away, prematurely and unexpectedly, before he could see its completion.}
\altaffiltext{$^\ddagger$}{The National Center for Atmospheric Research is sponsored by the National Science Foundation.}

\begin{abstract}
We realized a laboratory experiment to study the scattering polarization of the \ion{Na}{1} D-doublet at 589.0 and 589.6\,nm in the presence of a magnetic field. This work was stimulated 
by solar observations of that doublet, which have proven 
particularly challenging to explain through available models of polarized line formation, even to the point of casting doubts on our very understanding of the underlying physics. The purpose of the experiment was to test a quantum theory for the polarized scattering of spectrally flat incident radiation, on which much of the current magnetic diagnostics of stellar atmospheres is based. The experiment has confirmed the predictions of that theory, and its adequacy for the modeling of scattering polarization under flat-spectrum illumination.
\end{abstract}


\section{Introduction}
Over the past few decades, scattering polarization and its 
modification in the presence of a magnetic field
have become fundamental diagnostics of many physical properties 
of astrophysical plasmas \citep{review-1,review-2}. In particular, 
spectrally resolved observations of the polarized radiation from the 
solar disk near the limb, using high sensitivity ($\mathrm{S/N}\gtrsim 10^3$) 
instrumentation, have produced an extremely rich
amount of data (the so-called ``Second Solar Spectrum'' 
\citep{SS2-1,SS2-2}) of great diagnostic value \citep{diagnostic-1,diagnostic-2,diagnostic-3,diagnostic-4}. 
However, the interpretation of these observations has often proven to be difficult, and continues to challenge our understanding of how polarized radiation is produced and transported in the solar atmosphere.

One notable example is the linear polarization the D$_1$ resonance line of neutral sodium at 589.6\,nm, which has been the target of many observations \citep{D1obs0-1,D1obs0-2,D1obs1-1,D1obs1-2}. In the optically thin limit, this $J\,{=}\,1/2\leftrightarrow J'\,{=}\,1/2$ transition cannot produce broadband linear polarization, despite the polarizability of its hyperfine-structure (HFS) levels \citep{expectedD1-1,expectedD1-2,expectedD1-3}. This is because the spectral shape of its emissivity turns out to be anti-symmetric, and so it averages out to zero 
when the transition is spectrally unresolved. However, observations by \cite{D1obs0-1} and \cite{D1obs0-2} surprisingly had shown the presence of a strong linear polarization signal in the line core, raising many questions about its origin, and even on the reliability of those observations \citep{D1obs1-1,D1obs1-2}. While the complexity 
of the line-formation problem in the optically thick and magnetized 
atmosphere of the Sun is expected to play a role in determining the 
spectral shape of this line, the ``enigma'' posed by those observations has even brought some authors \citep{doubts-1,doubts-2} to questioning the adequacy of the quantum-electrodynamic formalism on which many of our interpretation tools for solar polarimetric observations are based \citep{LL04}. 
This impasse convinced us of the need to put this theoretical framework to the test with a specifically designed laboratory experiment.

\begin{figure}
\centering
\includegraphics[width=.49\textwidth]{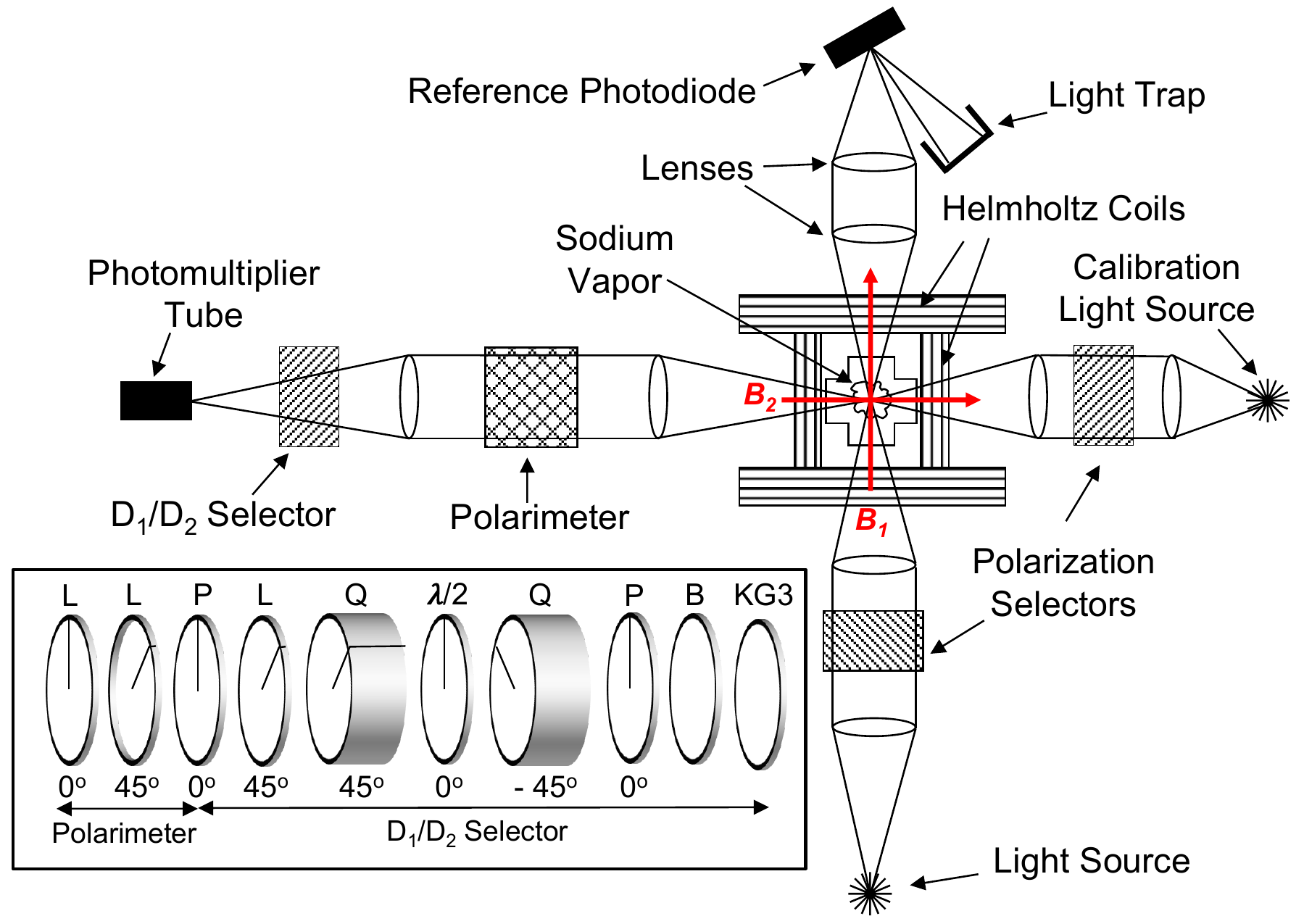}
\caption{Top-view diagram of the experimental setup. The four ``legs'' 
of the experiment are (clockwise from the bottom): input beam, scattered 
light analysis, light-level monitor, and calibration. The inset shows the elements of the polarimeter and D$_1$/D$_2$ selector (L\,=\,LCVR, P\,=\,polarizer, Q\,=\,quartz plate, $\lambda/2$\,=\,half-waveplate, B\,=\,9.5\,nm blocker). }
\label{fig:setup}
\end{figure}

\section{Experiment}
\subsection{Experimental Setup}
We built a scattering experiment where a vapor of neutral sodium under controlled conditions of temperature and magnetic field is illuminated by a light beam. The scattered radiation is analyzed polarimetrically, separately for the D$_1$ (3p $^2$P$_{1/2}$ $\rightarrow$ 3s $^2$S$_{1/2}$, 589.6\,nm) and D$_2$ (3p $^2$P$_{3/2}$ $\rightarrow$ 3s $^2$S$_{1/2}$, 589.0\,nm) transitions.

A top-view schematics of the experiment is shown in Figure~\ref{fig:setup}. 
This consists of a \ion{Na}{1} vapor cell surrounded by two 
air-cooled Helmholtz-coil pairs, and flanked by four ``legs'' with different functions. 
Light enters the apparatus from the bottom leg, is focalized at the center of the vapor cell, and the light scattered from 
the vapor at $90^\circ$ is analyzed in the left leg. The top leg uses a photodiode to monitor 
the light level of the source, and the right leg is used to input specific 
polarization states for the purpose of polarimetric calibration.
 
The center of the sodium cell is located at the intersection of the four legs 
of the apparatus. 
The sodium is evaporated into the cell from a reservoir 
which is temperature controlled at a typical value of 205 $^\circ$C. Along with the sodium vapor, the cell also contains $17\,$mmHg of Ar buffer gas.
The two Helmholtz-coil pairs allow the generation of a magnetic field 
between 0 and 150\,G with any desired direction in the scattering plane 
(plane of Figure~\ref{fig:setup}).

To ensure the condition of \emph{complete frequency redistribution} (CRD; see Modeling section) of the scattered 
radiation, we employed a 50\,W halogen bulb with stabilized output, 
which provides a largely flat and structureless spectrum over the frequency 
range of the D lines. 
%
%
An input polarization selector, consisting of a linear polarizer mounted 
in a precision rotation stage and a fixed $\lambda/4$ retarder, can be 
placed in the beam following the light source, allowing an arbitrary 
polarization state to be input to the vapor. In this letter, we
only present data and modeling for the case of unpolarized
input. 

The analysis leg consists of a Stokes polarimeter, a filter that 
selects the D line to be observed, and a photomultiplier tube (PMT) with a gain of approximately $2\times10^6$.
Details of the polarimeter and D$_1$/D$_2$ selector are shown in the inset diagram of Figure~\ref{fig:setup}.
The polarimeter consists of two Nematic Liquid Crystal Variable Retarders 
(LCVRs) followed by a linear polarizer. The LCVRs are 
oriented with their fast axes at $0^\circ$ and $45^\circ$, with the
linear polarizer also oriented at $0^\circ$. 
The orientation of this polarizer sets the reference direction of 
positive Stokes $Q$, which is approximately normal to the scattering plane. 
This system allows the analysis of the complete polarization 
state of the scattered light by measuring its intensity at 
selected retardations of the two LCVRs.

The D$_1$/D$_2$ line selector consists of a birefringent crystal between 
polarizers \citep{Ma74}, producing a channel spectrum with a free 
spectral range equal to twice the separation of the D doublet ($1.195\,$nm).
In order to minimize the shift of the bandpass with inclination angle 
through the selector, we have used quartz crystals in a wide-fielded 
configuration \citep{Ly44-1,Ly44-2}. 
The channel spectrum is shifted by a third Nematic LCVR with its 
fast axis aligned with that of the first crystal, which allows the 
electro-optical selection of either of the D lines. For simplicity, 
the analyzing polarizer of the polarimeter serves also as the entrance 
linear polarizer of the D$_1$/D$_2$ selector.
To limit the number of unwanted orders of the D$_1$/D$_2$ selector we additionally 
employ a 9.5\,nm wide interference filter centered 
at 590.5\,nm (blocker), and a Schott KG3 filter. To compensate for thermal shifts of the D$_1$/D$_2$
selector, we monitor its temperature and adjust the 
LCVR voltage to achieve the proper tuning.

The calibration leg contains a light source and 
input polarization selector identical to those in the input-beam leg. For the purpose of polarimetric calibration, light is 
input from the calibration leg into the analysis leg in the absence 
of sodium vapor (cold cell) and magnetic field. By measuring the output signal for
known input polarization states, we can compute the response matrix
of the polarimeter, 
which maps the measured Stokes vectors to the true ones.

\subsection{Measurements}
We measured the scattering polarization of the D lines 
in the presence of a magnetic field in the scattering plane, with strength between 0 and 150\,G 
in steps of 10\,G, and inclination from the direction of the incident 
radiation between $0^\circ$ and $90^\circ$ (respectively, $\bm{B}_1$ 
and $\bm{B}_2$ in Figure~\ref{fig:setup}) in steps of $30^\circ$. The 
calibration data were obtained before and after the scattering measurements. 
A measurement of the background signal was taken at the beginning of the experiment with the cold cell and no magnetic field. This background is 
a combination of Rayleigh scattering of the incident radiation by the 
Ar buffer gas, and parasitic reflections off the cell walls that make it into the analysis leg.

A computer running LabVIEW performs all experiment controls and 
data logging functions. The voltage output of the PMT is digitized 
with 16-bit precision. Each measurement consists of an average of
$10^4$ samples taken over $250\,$ms followed by a $125\,$ms delay to
allow for LCVR relaxation. A Stokes vector measurement is obtained 
by measuring the intensity coarsely corresponding to the six 
modulated states $I\pm(Q,U,V)$, and making the proper combinations 
and polarization cross-talk corrections to obtain $I,Q,U,V$. This is
accomplished by multiplying the measured Stokes vector by the 
polarimeter response matrix to obtain the true Stokes vector. 
The elements of the resulting Stokes vector have a typical uncertainty of $\sim10^{-3}$.

\section{Modeling}
To model the scattering polarization from the \ion{Na}{1} 
vapor we rely on several physical assumptions:

1) The flat spectrum of the light source implies
that radiation scattering can be described as the incoherent 
succession of single-photon absorption and re-emission \cite[CRD hypothesis;][]{CRD-1,CRD-2,LL04}.

2) Isotropic elastic collisions with the Ar buffer gas 
contribute to the statistical equilibrium of the \ion{Na}{1} atoms, 
leading to a partial depolarization of the atomic levels. The 
corresponding two depolarizing rates (respectively, for the 
orientation and the alignment of the atomic levels) are 
free parameters of the model. For simplicity, we adopt the same rates 
for the ground and excited states. However, the ensuing 
depolarization is nearly total for the ground state because of 
its much longer lifetime. 

3) In order to fit the data, we found necessary to add a small collisional de-excitation rate to the statistical equilibrium of the \ion{Na}{1} atoms. 
A possible explanation is that the collisions with the Ar buffer gas may not be perfectly elastic. However, the low temperature of the sodium vapor implies that collisional excitation from the ground state is negligible.
Thus the observed line intensity is dominated by the resonance scattering of the incident radiation, without any measurable contribution from Planckian 
radiation at the vapor temperature. 

Additionally, collisional transfer between the P$_{1/2}$ and P$_{3/2}$ levels can be important, as the energy separation is about $10^3$ times smaller than the excitation potential of the D-doublet. 
These transfer collisions predominantly produce a depolarization of the levels, adding to the effect of elastic collisions already considered. Since the relative contribution between transfer and elastic collisions to this depolarization is not constrained by our data, we chose to simply ignore transfer collisions in our model.

4) The gas cell is operated at a regime around unit optical depth.
Hence, differential saturation of the line components
must be taken into account \citep{satur-1,satur-2}. Additionally, polarization effects due to quantum 
interference between the fine-structure levels of 
the atom cannot in principle be ruled out under our experimental conditions. 
All these effects can confidently be modeled assuming that the fraction of the vapor contributing to the scattered radiation has spatially homogeneous thermodynamic and magnetic properties.

The differential saturation of the magnetic components of the lines \citep{satur-1,satur-2} turns out to be essential for the 
interpretation of the experimental results. In contrast, for the 
particular thermal and magnetic regimes of the experiment, our 
modeling shows that quantum interference between the 
P$_{1/2}$ and P$_{3/2}$ levels brings a much smaller, yet measurable, correction to the polarization.

\begin{figure*}
\includegraphics[width=1.0\textwidth]{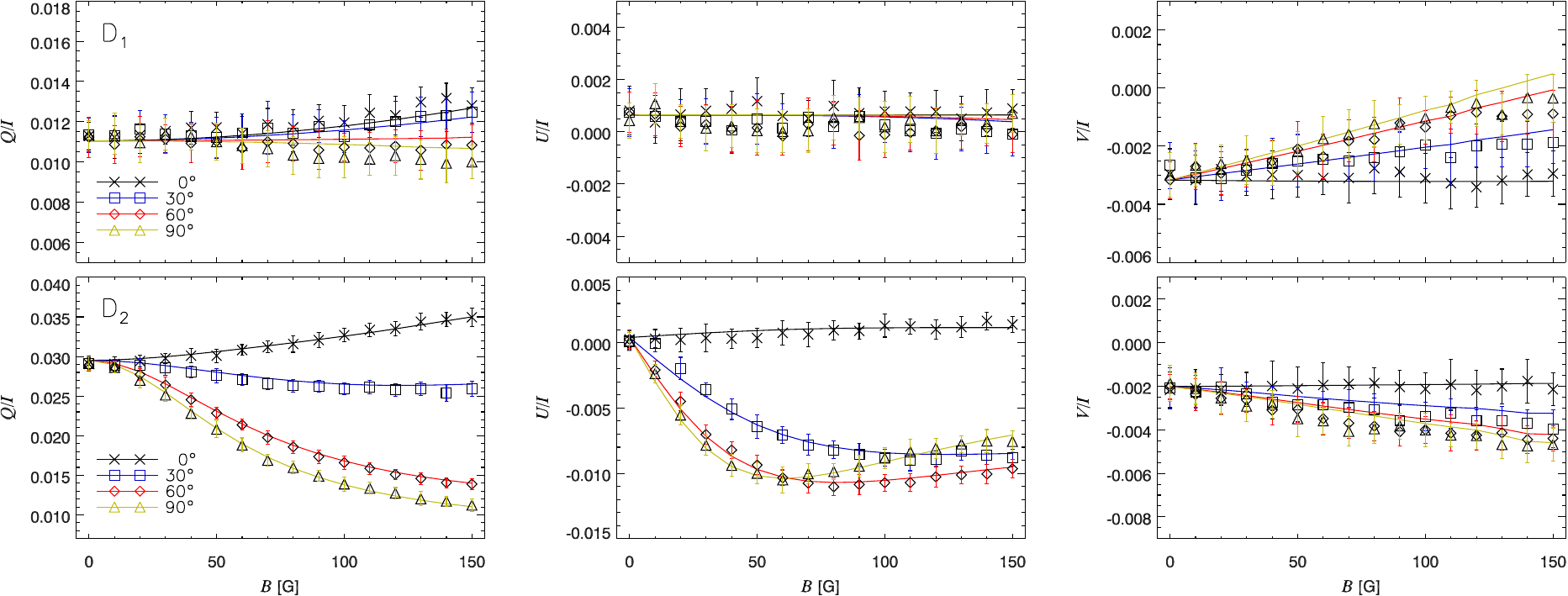}
\caption{Broadband fractional polarization $Q/I$, $U/I$, and $V/I$ (left to right) 
of the D$_1$ (top) and D$_2$ (bottom) lines as a function of magnetic field strength, 
for various geometries of the applied magnetic field. The measurements are
represented by different symbols (with error bars) and colors, for different values 
of $\vartheta_B$.
The continuous curves of matching color represent the model.} 
\label{fig:Stokes}
\end{figure*}

5) Magnetic-induced dichroism affects the transfer of both the sodium emission and the 
background radiation through the optically thick vapor. Hence, the background measurements cannot simply be subtracted from the experimental data, in order 
to isolate the contribution of the D lines to the observed polarization.
Instead, we must treat the background radiation as a boundary term in the solution of the radiative transfer equation for the Stokes vector $\bm{S}\equiv(I,Q,U,V)$,
\begin{equation} \label{eq:PRTE}
\frac{\mathrm{d}}{\mathrm{d} s}\,\bm{S} = -\mathbf{K}\,\bm{S}+\bm{\varepsilon}\;.
\end{equation}
where $s$ is the coordinate along the optical path, $\bm{\varepsilon}$ is the 
polarized emissivity vector (source term), and $\mathbf{K}$ is the 
$4\times 4$ absorption matrix, which also accounts for dichroism and magneto-optical 
effects \citep{LL04}.
The experimental data must then be compared with the numerical solution of 
eq.~(\ref{eq:PRTE}) in a spatially homogeneous medium \cite[][\S8.3]{LL04}, taking into account the background term, and after convolution with the transmission profile of the D$_1$/D$_2$ selector.

The above hypotheses suggest that we adopt the formalism of the \emph{multi-term} atom with HFS \citep{CM05-1,CM05-2} to model the scattering polarization by the magnetized sodium vapor, as this takes into account the effects of quantum interference between the P$_{1/2}$ and P$_{3/2}$ levels.
However, it is instructive to look at the algebraic formulation of the \emph{multi-level} atom with HFS given by \cite{LL04}, because it allows us to better grasp how the free parameters of the model and the magnetic field affect the scattering polarization in each of the two D lines.

The broadband polarized emissivity due to radiation scattering in a two-level atom 
$(J_\ell,J_u)$ with HFS is \cite[cf.][\S10.22]{LL04} 
\begin{equation} \label{eq:emiss}
\bar\varepsilon_i(\bm{\Omega})=k_{\rm L}^{\textrm{\tiny A}}
	\oint\frac{\textrm{d}\bm{\Omega}'}{4\pi}
	\sum_{j=0}^3 P_{ij}(\bm{\Omega},\bm{\Omega}';\bm{B})_{\rm hfs}\,
	S_j(\bm{\Omega}')\;,
\end{equation}
where $P_{ij}(\bm{\Omega},\bm{\Omega}';\bm{B})_{\rm hfs}$ is the
\emph{Hanle phase matrix}, and $i,j=0,1,2,3,$ enumerates the four 
Stokes parameters $I,Q,U,V$. The interpretation of eq.~(\ref{eq:emiss}) is
straightforward: the Stokes parameter $S_j(\bm{\Omega}')$
of the incident radiation along the direction $\bm{\Omega}'$
is scattered into the direction $\bm{\Omega}$ and 
polarization state $i$, with a frequency integrated cross-section given 
by the line absorption coefficient, $k_{\rm L}^{\textrm{\tiny A}}$. 
We recall that the incident radiation in our experiment has no spectral 
structure around the transition frequency of the line. 
Evidently, the assumption of a spectrally flat incident radiation is necessary in order to write
eq.~(\ref{eq:emiss}).

The Hanle phase matrix is given by
\begin{eqnarray} \label{eq:redistr}
P_{ij}(\bm{\Omega},\bm{\Omega}';\bm{B})_{\rm hfs}
	&=&\sum_{KK'Q} 
	(-1)^Q\,{\cal T}^K_Q(i,\bm{\Omega})\,
	{\cal T}^{K'}_{-Q}(j,\bm{\Omega}') \nonumber \\
&&\mathop{\times} W_{KK'Q}(J_\ell,J_u;\bm{B})_{\rm hfs}\;,
\end{eqnarray}
where the polarization tensors $T^K_Q(i,\bm{\Omega})$, with $K=0,1,2$ and
$Q=-K,\ldots,K$, characterize the scattering geometry, and are tabulated in \cite{LL04}. The \emph{line polarizability factor}
$W_{KK'Q}(J_\ell,J_u;\bm{B})_{\rm hfs}$ describes the magnetic dependence of the Hanle phase matrix. When stimulated emission and the polarization of the $J_\ell$ level can both be neglected, like in the case of our experiment, the polarizability factor can be expressed in algebraic form \cite[cf.][ eq.~(10.167)]{LL04}:
\begin{widetext}
\begin{eqnarray} \label{eq:polar}
W_{KK'Q}(J_\ell,J_u; \bm{B})_{\rm hfs}
&=& \frac{3(2J_u+1)}{2I+1}
	\sixj{1}{1}{K}{J_u}{J_u}{J_\ell}
	\sixj{1}{1}{K'}{J_u}{J_u}{J_\ell} \\
&&\kern -1.6in 
	\times 
	\sum_{F_u F_u' F_u'' F_u'''}
	\sqrt{(2K+1)(2K'+1)(2F_u+1)(2F_u'+1)(2F_u''+1)(2F_u'''+1)}\,
	\sixj{J_u}{J_u}{K}{F_u}{F_u'}{I}
	\sixj{J_u}{J_u}{K'}{F_u''}{F_u'''}{I}
		\nonumber \\
\noalign{\vskip -6pt}
&&\kern -1.6in 
	\times\sum_{f_u f_u'} \sum_{i j}
	C_{F_u}^i(J_u f_u)\,
	C_{F_u''}^i(J_u f_u)\,
	C_{F_u'}^j(J_u f_u')\,
	C_{F_u'''}^j(J_u f_u') 
	\threej{F_u}{F_u'}{K}{-f_u}{f_u'}{-Q}
	\threej{F_u''}{F_u'''}{K'}{-f_u}{f_u'}{-Q} 
	\nonumber \\
\noalign{\vskip -6pt}
&&\kern -1in 
	\times \left\{ 1+\delta\apx{$K$}_{J_u}+\epsilon_{J_uJ_\ell}
	+\mathrm{i}[\omega_j(J_uf_u')-\omega_i(J_u f_u)]/
		A_{J_uJ_\ell}\right\}^{-1}\;. \nonumber
\end{eqnarray}
\end{widetext}
The coefficients $C_{F_u}^i(J_u f_u)$, with $F_u=|J_u-I|,\ldots,J_u+I$, are the components of the {$i^\mathrm{th}$ eigenvector of the HFS subspace of $J_u$ with magnetic quantum number $f_u$, and $\omega_{i}(J_u f_u)$ is the corresponding eigenvalue. They are determined 
via diagonalization of the magnetic Hamiltonian, assuming the direction of $\bm{B}$ as the quantization axis.
In the denominator of eq.~(\ref{eq:polar}), the imaginary term accounts for polarization effects associated with the energy differences 
between the atomic eigenstates (Hanle effect, HFS depolarization, level-crossing interference). 
$\delta\apx{$K$}_{J_u}$ and $\epsilon_{J_uJ_\ell}$ are, respectively, the 
depolarizing and inelastic collision rates, expressed in units of the 
Einstein coefficient $A_{J_uJ_\ell} \approx 6.2{\times}10^8\,\rm s^{-1}$ \cite[cf.][eq.~(10.54)]{LL04}. 
For the contribution of the level population to the emissivity ($K=0$), $\delta\apx{0}_{J_u}=0$, thus
the polarizability factor only contains the free parameters
$\delta\apx{1}_{J_u}$ (orientation relaxation), $\delta\apx{2}_{J_u}$ 
(alignment relaxation), and $\epsilon_{J_uJ_\ell}$ (collisional de-excitation).

For the multi-level atom, a distinct set of these three free parameters must 
be specified for each of the two D lines. 
%
In the multi-term formalism, 
instead, we only need the three parameters $\delta\apx{1,2}\equiv\delta\apx{1,2}_{L_u}$ 
and $\epsilon\equiv\epsilon_{L_u L_\ell}$, expressed in units of the D-doublet spontaneous rate $A_{L_u L_\ell}\approx 6.2{\times}10^8\,\rm s^{-1}$, where $L_u=1$ and $L_\ell=0$.
On the other hand, 
 for the multi-term atom, 
 an algebraic expression of the broadband emissivity 
 analogous to eqs.~(\ref{eq:emiss})--(\ref{eq:polar}) cannot be attained \emph{separately} for 
 each line of the doublet.

\section{Results} 
Figure~\ref{fig:Stokes} reports one set of measurements (resulting from the average of 12 different realizations of the experiment) 
of the broadband fractional polarization of the two D lines (symbols with error bars). In Figure ~\ref{fig:Stokes}, the continuous 
curves represent the fit of the experimental data provided by the model described in the previous 
section. It is important to remark that the zero-field values in all plots, except for the $Q/I$ polarization of D$_2$, are dominated by the transfer of the background polarization through the optically thick vapor. In the absence of background radiation, those values would be zero (within the polarimetric accuracy of the experiment). The intensity and polarization of the background are measured at the beginning of the experiment (cold cell). The ratio $I_\mathrm{bkg}/(I_\mathrm{line}+I_\mathrm{bkg})$ turns out to be about 17\% for D1 and 
12\% for D2, while the polarization of the background is very consistent between the two spectral ranges, with 
($Q_\textrm{bkg},U_\textrm{bkg},V_\textrm{bkg})/I_\mathrm{bkg}\simeq(1,0.064,0.004,-0.018)$.

Numerical modeling based on eqs.~(\ref{eq:emiss})--(\ref{eq:polar}) predicts that all states of polarization of D$_1$, as well as the $V/I$ polarization of D$_2$ should remain largely insensitive to the magnetic field in an optically 
thin vapor, well below the $10^{-3}$ sensitivity level of our experiment.
The large departures from 
such ideal behavior observed in the experimental data, especially for the $V/I$ polarization, are mainly 
due to the differential saturation of the magnetic components of the lines as they are transferred through 
the optically thick vapor \citep{satur-1,satur-2}. In order to fit the measurements, we determined an optical depth 
$\tau_\mathrm{D_2}\approx 1.3$. The non-flat behavior of the $U/I$ polarization of D$_2$ 
for $\vartheta_B=0^\circ$ is explained by a small error of the apparatus in setting the desired magnetic field 
inclination, which we modeled with a $-2^\circ$ offset from the nominal values of $\vartheta_B$.
 
The remaining free parameters of the model are the depolarizing collision rates 
$\delta\apx{1,2}$ and the de-excitation collision rate $\epsilon$. The value of $\delta\apx{2}$ 
strongly affects the linear polarization amplitudes of D$_2$, and characteristically the location 
of the two crossing points among the $U/I$ polarization curves for 
$\vartheta_B\ne 0$. We used these constraints to determine a value 
$\delta\apx{2}\approx 19$.
The $\delta\apx{1}$ rate affects instead the  
$V/I$ polarization caused by the presence of atomic orientation. In the case of unpolarized input, this contribution is rapidly suppressed by depolarizing
collisions. Thus, the value of $\delta\apx{1}$ is only weakly constrained by the data shown in Figure~\ref{fig:Stokes}. However, when the incident light is circularly polarized, the observed $V/I$ signals are much larger (by a factor ${\sim} 10$, in the case of D$_2$) than those shown in Figure~\ref{fig:Stokes}. Using such measurements (not reported here), we could determine $\delta\apx{1}\approx 13$.
%
Finally, by matching the zero-field value of the $Q/I$ polarization of D$_2$, after taking into account the 
depolarization produced by $\delta\apx{2}$, we estimated $\epsilon\approx 0.44$.

\section{Conclusions}
The agreement between theory and experiment shown in Figure~\ref{fig:Stokes} is remarkable, considering that the fitting of the reported data (384 independent polarization measurements) practically relies on only three model parameters, 
$\tau$, $\delta\apx{2}$, and $\epsilon$. This demonstrates that the quantum-electrodynamic formalism on which our model of scattering polarization in the CRD limit is based \citep{LL04} is completely adequate when the incident radiation is spectrally flat over the wavelength range of the atomic transition.

The \ion{Na}{1} D lines, however, are among the strongest absorption features of the solar spectrum, and the flat-spectrum approximation breaks down in the solar atmosphere, especially with regard to the treatment of the quantum interference between the P$_{1/2}$ and P$_{3/2}$ levels. Therefore, new polarization effects due to the partial redistribution of the radiation frequency (PRD) can be expected for these lines \citep{modeling-1,modeling-2}.
Recent work \citep{PRD-5,PRD-1,PRD-2,PRD-4,PRD-3} has formally extended the theory of \cite{LL04} beyond the CRD limit, in order to model PRD effects in radiation scattering. Indeed, when these effects are taken into account in the modeling of the polarized D$_1$ line \citep{modeling-1,modeling-2}, even its finer spectral details as observed on the Sun \citep{D1obs1-1,D1obs1-2} can be reproduced.

The successful interpretation of our experiment provides compelling evidence of the fundamental validity of the quantum-electrodynamic formalism used to interpret the many polarization phenomena routinely observed on the Sun and in other astrophysical objects. At the same time, together with the recent modeling by \cite{modeling-1} and \cite{modeling-2}, our results strongly support the conclusion that the peculiarities of the observed polarization of the D$_1$ line \citep{D1obs0-1,D1obs0-2,D1obs1-1,D1obs1-2} must be traced back to the complexity of the line formation problem in realistic atmospheric scenarios, or in extreme cases to possible instrumental effects that must be identified and corrected for.

\section{acknowledgments}
Financial support for this experiment was provided by the 
National Center for Atmospheric Research through the Director's 
Opportunity Funds. We thank G. Card for his contribution to the design and construction of the experiment. The authors have benefited from many discussions with 
several colleagues, who at times have also assisted in various aspects of the experiment. 
In particular, we thank A.~de Wijn, R.~Manso Sainz, A.~L\'opez Ariste, and J.~O.~Stenflo. We thank J.~Trujillo Bueno for helpful comments and suggestions on the final version of the manuscript. 

\end{document}